# Current-voltage characteristics in Ag/Au nanostructures at resistive transitions

Saurav Islam[1#], Rekha Mahadevu[2#], Subham Kumar Saha[2], Phanibhusan Singha Mahapatra[1], Biswajit Bhattacharyya[2], Dev Kumar Thapa[1], T. Phanindra Sai[1], Satish Patil[2], Anshu Pandey[2], and Arindam Ghosh[1,3]

**Affiliations:**

[1]Department of Physics, Indian Institute of Science, Bangalore 560012, India

[2]Solid State and Structural Chemistry Unit, Indian Institute of Science, Bangalore 560012, India

[3]Center for Nano Science and Engineering, Indian Institute of Science, Bangalore 560012, India

*Correspondence to: arindam@iisc.ac.in

#Contributed equally

**Transitions to immeasurably small electrical resistance in thin films of Ag/Au nanostructure-based films have generated significant interest because such transitions can occur even at ambient temperature and pressure. While the zero-bias resistance and magnetic transition in these films have been reported recently, the non-equilibrium current-voltage ($I-V$) transport characteristics at the transition remains unexplored. Here we report the $I-V$ characteristics at zero magnetic field of a prototypical Ag/Au nanocluster film close to its resistivity transition at the critical temperature $T_c$ of $\approx 160$ K. The $I-V$ characteristics become strongly hysteretic close to the transition and exhibit a temperature-dependent critical current scale beyond which the resistance increases rapidly. Intriguingly, the non-equilibrium transport regime consists of a series of nearly equispaced resistance steps when the drive current exceeds the critical current. We have discussed the similarity of these observations with resistive transitions in ultra-thin superconducting wires via phase slip centres.**

The observation of resistive transition in Ag/Au nanostructure-based films has recently triggered a widespread excitement and debate [1]. The transition to resistance below $\sim$ few μΩ, that corresponds to electrical resistivity $< 10^{-12}$ Ω-m at temperatures as high as $\sim 286$ K and ambient pressure, have been attributed to multiple possibilities, including a superconductor-like macroscopic order [2], as well as percolative decoupling of the current and voltage paths [3]. While concurrent diamagnetic transition possibly indicates emergent many-body coherence, new experimental characterization of the transition is evidently required. Nonequilibrium electrical transport is a powerful tool to characterize many-body states where the existence of critical field [4–6], low-bias anomalies in differential conductance [7, 8], hysteresis [9] etc. carry distinctive signatures of the underlying electronic phase. Here, we have carried out detailed measurements of two and four-probe current-voltage ($I-V$) characteristics in a prototypical Ag/Au nanocluster film that exhibits the resistive transition at $T_c \approx 160$ K (at zero magnetic field). We find that the $I-V$ characteristics is strongly hysteretic close to the transition (within a temperature range of $\approx 50$ mK) and exhibits current-driven nonlinearity through discretely varying resistance states. Similar observations in other devices (Supplementary Information) indicates a generic behaviour that resembles phase-slip mediated nonlinearity and discreteness in one dimensional superconductors.

The preparation details of the Ag/Au nanostructure can be found in Ref [1]. Briefly, the process involves embedding ∼ 1 nm Ag nanoparticles into a gold matrix using colloidal methods. Ligands are subsequently removed during the film preparation stage to allow for electrical scontinuity. The film of average thickness ≈ 100 nm was deposited on a prefabricated multi-lead architecture in Van der Pauw geometry on a glass slide. The leads were made by thermal evaporation of 5/50 nm Cr/Au directly on the slide. Fig. 1a shows the typical optical micrograph of a sample used in our measurements. The AFM image of an identically prepared film over a $700^2$ nm area is shown in Fig. 1b. Fig. 1c shows the line scan from three different regions of the film. Averaging of multiple such scans from different regions indicates an overall surface roughness of ∼ 30 nm, with over ∼ 90 % spatial coverage. The four-probe resistance ($R$) across the device (e.g $I$ : 2,3 and $V$ : 6,7, see inset of Fig. 1d) is ≈ 0.5 Ω, which corresponds to specific resistivity ≈ $5 \times 10^{-8}$ Ω-m, which is a factor of 2 larger than that of bulk Au. $R$ is usually anisotropic, and varies by over a factor of ∼ ten between different parts of the film (Section I in Supplementary Information), likely due to spatial variation in the nanoparticle density and structural morphology of the film itself. To study the dependence of $R$ on temperature ($T$) we usually choose the set of contacts that display the least $R$ in the four-probe combination across the device. In all cases, irrespective of temperature $T$, the value of $R$ remained unchanged if the current and the voltage lead pairs were swapped, thereby ensuring Onsager reciprocity.

Fig. 1e shows the $T$-dependence of four-probe $R$ in two thermal cycles, denoted as thermal cycle 2 and 4, where $T$ was varied at ∼ 13 mK/min during both heating and cooling. Leads (2,3) and (6,7) were used for $I$ sourcing and $V$ measurement, respectively (inset of Fig. 1d). The value of $R$ is calculated from $V$ measured at $I = 1$ mA. The $R$ vs $T$ during all thermal cycles for this device can be found in Section II in the Supplementary Information. The transitions in $R$ are evident in both cycles, where the low resistance state $R$ ∼ 10 μΩ (Section III in Supplementary Information), corresponds to resistivity $10^{-12}$ Ω-m. To ensure electrical inter-connectivity between all leads in the low $R$ state, we measured the two-probe $I - V$ characteristics in pair-wise manner at $T = 160$ K immediately after thermal cycle 4. As seen in Fig. 1d, all $I - V$s are linear and varies between ≈ 5 Ω to ≈ 100 Ω. (We point out that the two-probe resistance will include contribution from the lead-film contact, thin evaporated leads, as well as the contact between the lead and the spring-loaded pin.) The metallic resistance and non-hysteretic linear $I - V$ dispel the concerns of a percolative decoupling between the current flow path and voltage leads [3]. This further confirms the well compacted nature of our nano-structured films, because weak coupling between grains is known to manifest in Coulomb blockade-induced non-linear transport and variable range hopping [10, 11]. We note: (1) A strong hysteresis between the heating and the cooling can make the transition temperature $T_c$ differ by nearly ≈ 5 K, within the same thermal cycle. (2) While $T_c$ changes from one cycle to the other, as seen in nearly 15 K difference between cycle 2 and 4, in general it increases to higher $T$ with ageing [Section II Supplementary Information] [1]. (3) The cooling step in cycle

2 has a re-entrant behaviour around 172 K that was observed before [1], as well as few other thermal cycles in this device (Section IV in Supplementary Information).

The four-probe $I-V$ characteristics were recorded during thermal cycle 4. The temperature was first stabilized at 160 K, following which we commenced slow heating at $\sim 13$ mK/min, while ramping $I$ between leads 2 and 3, and recording $V$ between 6 and 7. First $I$ was ramped from 0 to $+10$ mA, then decreased from $+10$ mA to $-10$ mA and finally returned to 0. Each such cycle lasted $\approx 120$ sec, during which $T$ changed by about $\approx 25$ mK. Fig. 2a and b show the evolution of the $I-V$ characteristics as a function of $T$ during the heating and cooling cycles, respectively. It is evident that the $I-V$ characteristics become hysteretic at the onset of the finite $R$ during heating (Fig. 2a) and also close to the drop in $R$ below the measurable limit during the cooling cycle (Fig. 2b). At higher temperature, the $I-V$s are non-hysteretic and expectedly linear. Notably, there is no discernible non-linearity in $I$ or differential resistance $dV/dI$ (See Section V in Supplementary Information), close to zero bias ($V \to 0$), that are often seen in metallic point contacts and break junctions [12, 13], or ballistic channels in high mobility electron gas in semiconductor [14, 15]. Importantly, the non-linearity in the increasing positive current segment (i.e. $I: 0 \to +10$ mA) in the $I-V$ shows upturn in the voltage drop at a current which increases as $T$ is lowered. This behaviour is clearer for the cooling cycle and shown separately in Fig. 2c. The characteristic current scale ($I_c$) can be evaluated by extrapolating from the large $V$ range (dashed lines in Fig. 2c), and found to increase rapidly from $\approx 0.7$ mA at $T = 161.42$ K to $\approx 9.5$ mA at $T = 161.36$ K.

In Figs. 2b and 2c, we also find that $V$ increases in abrupt jumps for $I > I_c$ (for increasing $I$), and follows a similar pattern while decreasing as $I$ returns to zero. Similar behavior was also observed in the negative current segment (Section VI in Supplementary Information), as well as in the heating cycle (Section VII in Supplementary Information). To explore this further, we have plotted the positive current segment of the $I-V$ characteristics at 161.36 K (Fig. 2b) in Fig. 3a. While $I-V$ remains largely featureless as $I$ is increased to $I_c$ (black vertical arrow), the piece-wise linear decrease in $V$ when $I$ decreases is evident. The red dotted lines converge to $(I, V): (0,0)$, indicating that the decrease in $V$ is governed by sequential removal of discrete resistive elements as $I$ decreases. This is shown in Fig. 3b, where we plot $R (= V/I)$, as a function of $I$. In addition to the stair-like decrease, we find the steps in resistance ($\Delta R$) are astonishingly similar in magnitude. We find $\Delta R \approx 0.030$ $\Omega$ between most resistance steps in Fig. 3b (blue dashed horizontal lines), but reduces by about half to $\Delta R \approx 0.015$ $\Omega$ at the highest current. Intriguingly, the resistance steps continue to remain roughly equispaced at $\Delta R \approx 0.015$ $\Omega$ for $I-V$ recorded at $T = 161.39$ K, as shown in Fig. 3c (also see Section VIII in Supplementary Information). Similar stair-like pattern and hysteresis have been observed in other samples as well with step sizes of similar magnitude (See Section IX in Supplementary Information).

Electric field-driven hysteresis in $I-V$ characteristics, and multi-state resistance switching phenomenon, are often encountered in solid-state memories, nano-structured films and, in general, percolative/strongly localized systems [16–18]. These, however, are important at large resistances ($R \sim 10^3$–$10^9$ $\Omega$), and unlikely to consist of precisely equal changes in $R$. In fact, the only process known to host such effects close to vanishingly small electrical resistance is the superconducting transition in ultra-narrow superconductors through the formation of phase slip centers (PSC) [4, 19–27]. PSCs act as localized resistive centers along the length of a superconducting filament, where the order parameter $|\psi|$ shrinks to zero, while the phase changes by $2\pi$. The resulting momentum relaxation supports discontinuous change of the time average quasi-particle chemical potential across the PSC. Assuming a filamentary flow of super-current in our films close to the transition (see schematic in Fig. 3d), it is possible that each step in the observed $I - V$ characteristics indicates the appearance (when $I$ increases beyond $I_c$) or disappearance (when $I$ decreases) of PSCs at randomly located weak links. Each PSC is expected to contribute equally to the differential resistance [4, 20], which may explain the uniformity of $\Delta R$ between the steps, however the factor of two difference in the observed values of $\Delta R$ will require further understanding. The pronounced hysteresis in the presence of PSCs is also well-studied, and attributed to Joule heating due to localized dissipation, although intrinsic sources of hysteresis cannot be completely ruled out [4, 20, 25].

For a quantitative analysis, we fitted the $T$-dependence of $R$ at the transition with the thermally activated phase slip (TAPS) model developed by Langer-Ambegaokar-McCumber-Halperin (LAMH) [22, 23] and Little [20]. Within this framework, resistance $R(T) = R_N \exp(-\Delta F/k_B T)$, is determined by thermal activation over the free-energy barrier, $\Delta F \approx k_B T_c R_Q A(1-T/T_c)^{3/2}/\rho_N \xi$, where $R_N$, $R_Q$ (= 6.45 k$\Omega$), $A$, $\rho_N$, and $\xi$ are normal state resistance, superconducting quantum resistance, effective cross-sectional area, normal state resistivity, and coherence length, respectively [24–26]. For fitting, we chose the $R - T$ trace in the heating stage of thermal cycle 2 in Fig. 1e because of the availability of larger number of data points at low $R$. The fit (solid blue line) in Fig. 4a, with two parameters $\alpha$ and $T_c$, appears to be satisfactory over about three decades in $R$, and yields $\alpha = R_Q A/\rho_N \xi \sqrt{T_c} \approx 7 \times 10^4$ K$^{-1/2}$ and $T_c \approx 178.8$ K. Although the phase slip occurs over the coherence length $\xi$, the voltage drop across the PSC is determined by the quasiparticle diffusion over a characteristic charge imbalance length $\Lambda_Q$, and leads to resistive contribution of $\Delta R = 2\Lambda_Q \rho_N/A$ for every PSC [20]. Taking $\Delta R \approx 0.015$ $\Omega$ from Fig. 3b and c, and combining with $\alpha$ from the fit, we find, $\Lambda_Q/\xi \approx 1.1$, suggesting the normal and superconducting electrochemical potentials vary over the similar spatial scale across the PSC. A quantitative estimation of $\xi$ is however difficult in the absence of a known scale of the filament cross-section. An estimate can be obtained by assuming this to be close to the height fluctuations $\sim 30$ nm in surface topography (Fig. 1c) (similar scale was also observed for nanoparticle agglomerates in scanning transmission electron microscopy). This yields $\xi \sim 1$ nm for $\rho_N \sim 10^{-8}$ $\Omega$-m, which is consistent with large $H_{c2}$ reported earlier in this system [1].

Finally, we make two further observations: (1) First, from $T_c$ estimated from TAPS fit, we find $I_c \propto (T_c - T)$, within experimental uncertainty, as shown in the inset of Fig. 2c. Similar $T$-dependence of critical current has been predicted within the Ambegaokar-Baratoff framework for tunnel-coupled granular superconductors [28, 29] that differs from the expectation of the Gizburg-Landau theory, where $I_c \propto (T_c - T)^{3/2}$ (dotted line in the inset of Fig. 2c). [4] (2) Second, by measuring the two-probe resistance ($R_{2P}$) between the current leads simultaneously with the four-probe $R$ across the transition, we find a small increase in $R_{2P}$ at the transition in all our devices (Fig. 4b). The increase in $R_{2P}$ is sample specific, varying from $\sim 0.5\ \Omega$ to $\sim 10\ \Omega$. While a microscopic understanding of this increase in "contact resistance" at the transition remains elusive, possible role of emergent quasi-particle gap at the lead-film interface cannot be ruled out [30].

In conclusion, we report detailed current-voltage ($I - V$) characteristics at the resistivity transition in thin films made from Ag/Au nanostructures. The $I - V$ characteristics show pronounced hysteresis and stair-like pattern as the resistance drops below the experimentally measurable limit at the transition. We also identify a temperature-dependent critical current scale, beyond which the resistance increases rapidly. The transition can be quantitatively analyzed with phase-slip events in quasi-one-dimensional flow of supercurrent, supporting possible formation of superconducting order.

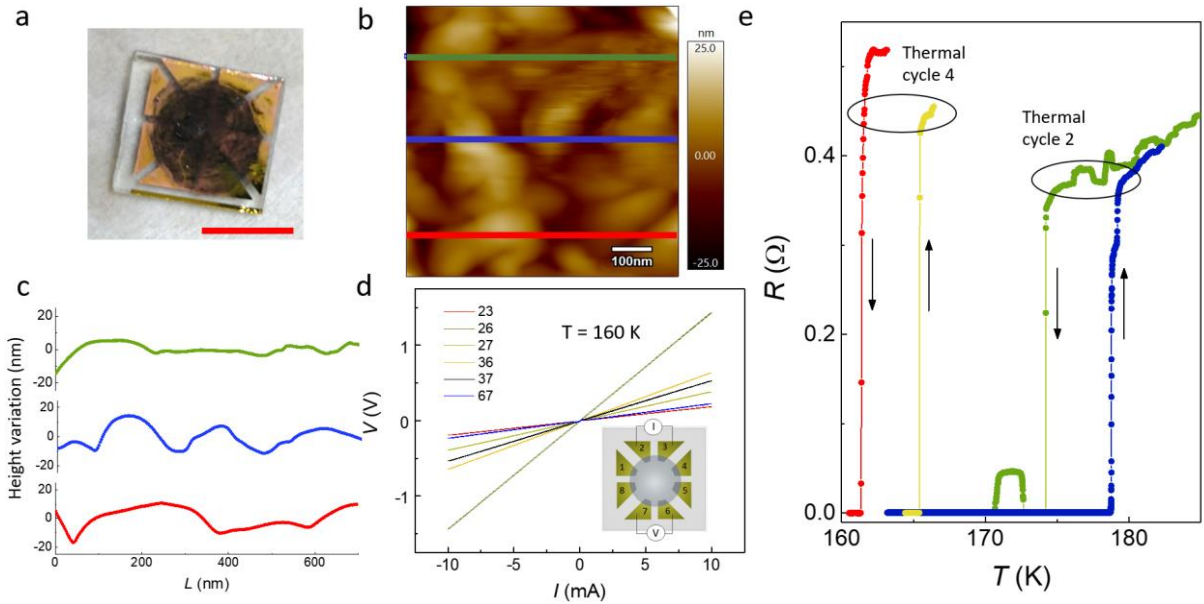

**Fig. 1. Structure and resistive transition in Ag/Au nanostructured film:** (a) Optical micrograph of a typical sample. The scalar bar is 5 mm. (b) Atomic force micrograph of the Ag/Au nanostructure film. (c) The height profile along three lines shown in (b). We find the average height variation of ∼ 30 nm in our films. (d) Two-probe current-voltage ($I - V$) characteristics between different contact configurations, displaying linear behavior, thus ensuring electrical continuity. The $I - V$ sweeps contain both forward and reverse sweeps of the driving current. (e) Resistance vs temperature displaying resistive transitions in two of the thermal cycles denoted as 2 and 4. Also see Section SII of Supplementary Information.

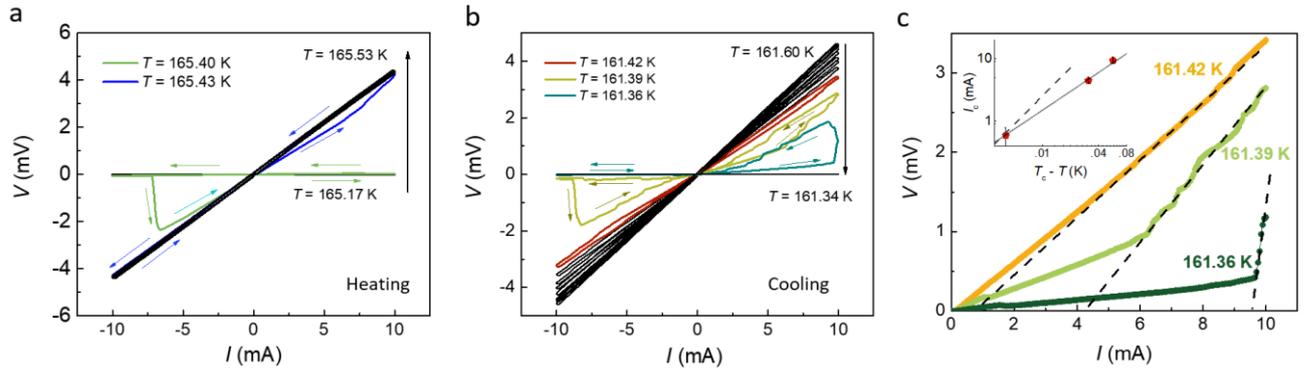

**Fig. 2. Current-voltage characteristics:** Four-probe $I-V$ characteristics of the sample at different temperatures through the resistive transition during the heating cycle (a) and cooling cycle (b). (c) The $I-V$ characteristics in the positive increasing segment of $I$ for the cooling cycle. The critical current ($I_c$) at which the film becomes resistive is identified by the extrapolation of the dashed lines to the current axis. Inset: Dependence of $I_c$ on temperature. The solid line indicates $I_c \propto (T_c - T)^\beta$, with $\beta = 1$. The dashed line corresponds to $\beta = 3/2$. $T_c$ is the transition temperature (See text).

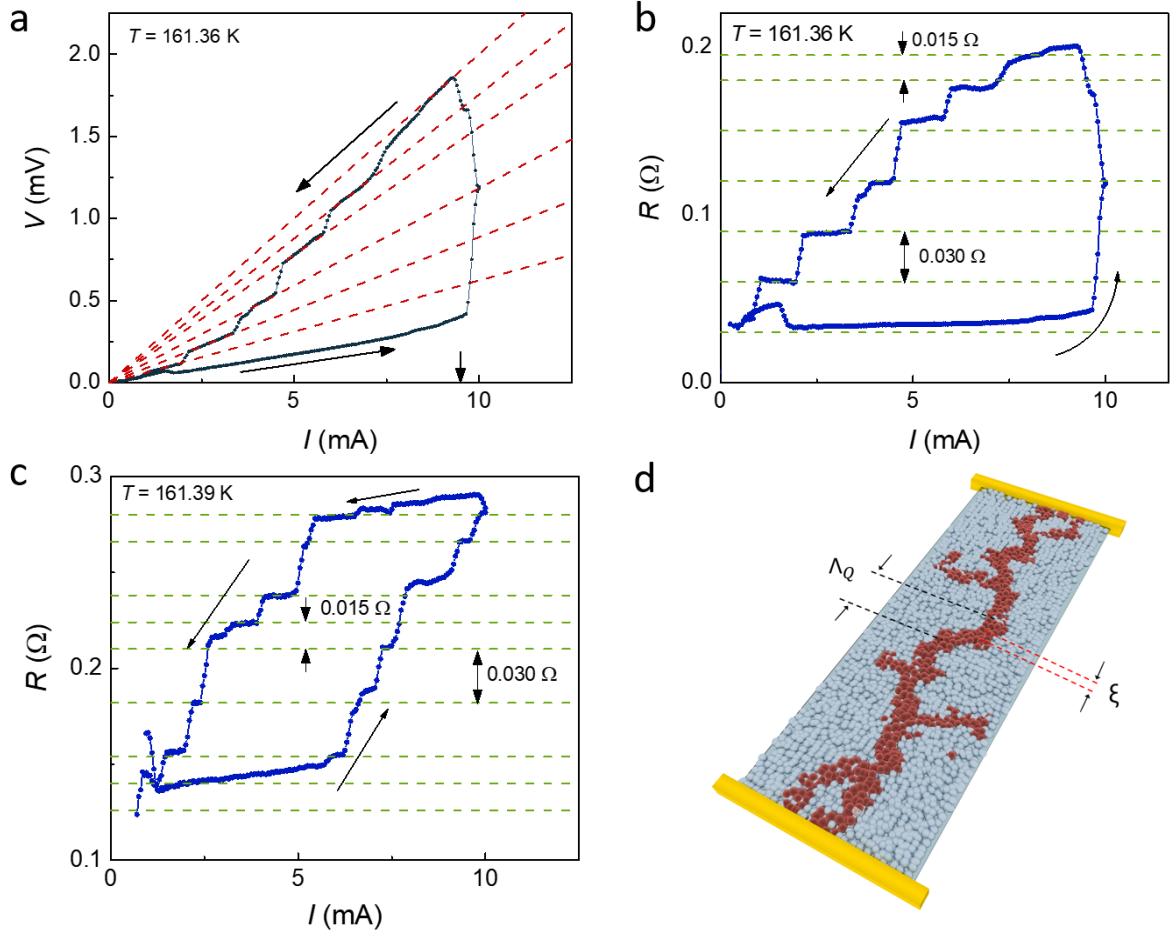

**Fig. 3. Stair-like hysteresis in non-equilibrium transport:** (a) The positive current segment of the $I-V$ characteristics at $T = 161.36$ K, obtained when the sample was cooled. (b), (c) $R (= V/I)$ as a function of $I$, obtained from $I-V$ data at (b) $T = 161.36$ K, and (c) $T = 161.39$ K, showing uniformly-spaced step-like features. (d) Schematic representation of phase slip centers for filamentary flow of supercurrent. $\Lambda_Q$ and $\xi$ are the charge imbalance length and coherence length, respectively.

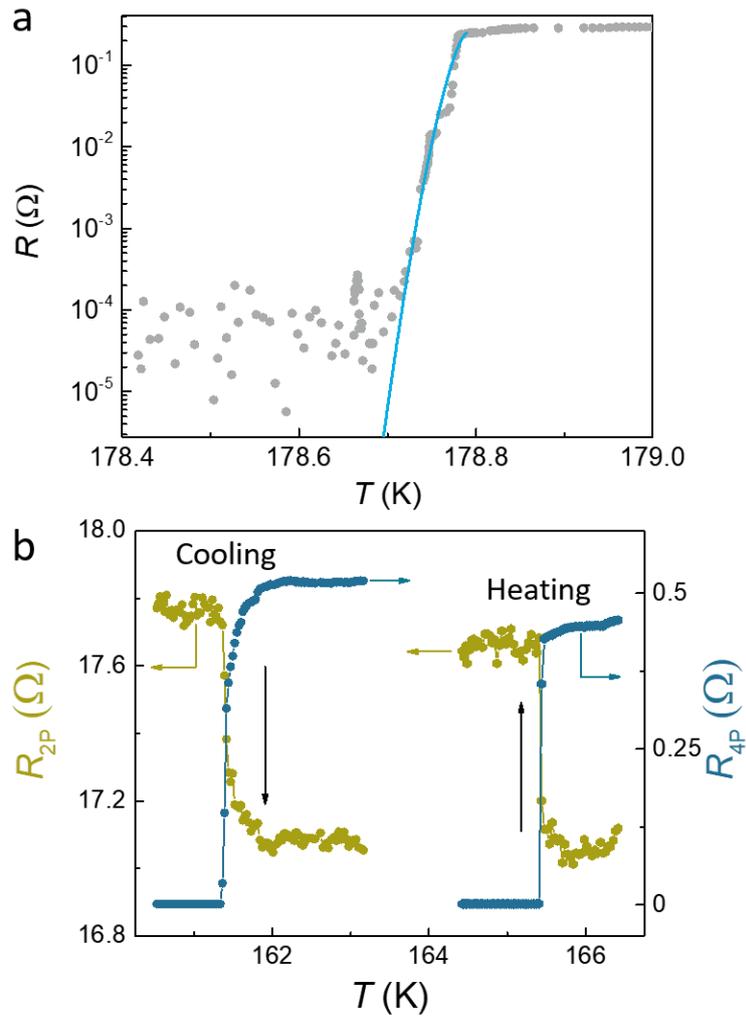

**Fig. 4. Temperature-dependence of the resistive transition and contact resistance:** (a) The fit (blue line) to the $R$ $vs$ $T$ data with the thermally activated phase slip model. The solid line corresponds to $R(T) \propto \exp[-\alpha(T_c - T)^{3/2}/T]$, where $T_c$ and $\alpha$ are fit parameters. (b) Correlation between two- and four-probe resistances during the resistive transition. An increase in the two-probe resistance across the transition can be observed.

# Current-voltage characteristics in Ag/Au nanostructures at resistive transitions

Saurav Islam[1], Rekha Mahadevu[2], Subham Kumar Saha[2], Phanibhusan Singha Mahapatra[1], Biswajit Bhattacharyya[2], Dev Kumar Thapa[2], T. Phanindra Sai[1], Anshu Pandey[2], and Arindam Ghosh[1,3]

[1]Department of Physics, Indian Institute of Science, Bangalore 560012

[2]Solid State and Structural Chemistry Unit, Indian Institute of Science, Bangalore 560012

[3]Center for Nano Science and Engineering, Indian Institute of Science, Bangalore 560012

Email: arindam@iisc.ac.in

**Section I: Anisotropy in resistance**

**Table I**

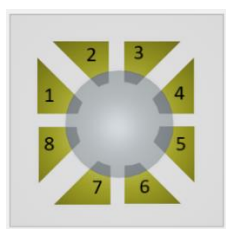

| Current probes | Voltage probes | Resistance (four probe) |
|---|---|---|
| 25 | 34 | 6.96 Ω |
| 36 | 45 | 4.55 Ω |
| 47 | 56 | 5.55 Ω |
| 58 | 67 | 4.3 Ω |
| 23 | 56 | 830 mΩ |
| 23 | 67 | 507 mΩ |
| 34 | 87 | 1.105 Ω |
| 26 | 87 | 3.2 Ω |

Table I: (Left) Schematic of the lead configuration. (Right) Four probe resistances measured in various contact configurations.

**Table II**

| Contacts | Resistance (two probe) |
|---|---|
| 12 | NA |
| 23 | 0.6 Ω |
| 34 | 6.5 Ω |
| 45 | 5.9 Ω |
| 56 | 24.2 Ω |
| 67 | 22.6 Ω |
| 78 | 3 Ω |
| 81 | NA |

Table II: Two probe resistance between different contact combinations, before cooling the sample, measured at room temperature. Combinations where electrical continuity was absent is referred to as NA.

**Section II: Effect of thermal cycles**

The resistance vs temperature data in all thermal cycles recorded in the device are shown in Fig. S1. The $T_c$ (= 120 K) is lowest in the maiden thermal cycle, below which the $R$ drops to about 10 µΩ. The $T_c$ varies in subsequent thermal cycles, while in the final cycle (5) the $T_c$ is as high as 250 K. The later observation has been reported before and likely due to ageing [1]. We encountered a residual resistance of ~ 1.6 mΩ in the low $R$ state in later thermal cycles that was manually subtracted from subsequent transport data. The resistance during thermal cycle 4 was extracted from the voltage drop at $I = 1\text{mA}$ in the current voltage characteristics.

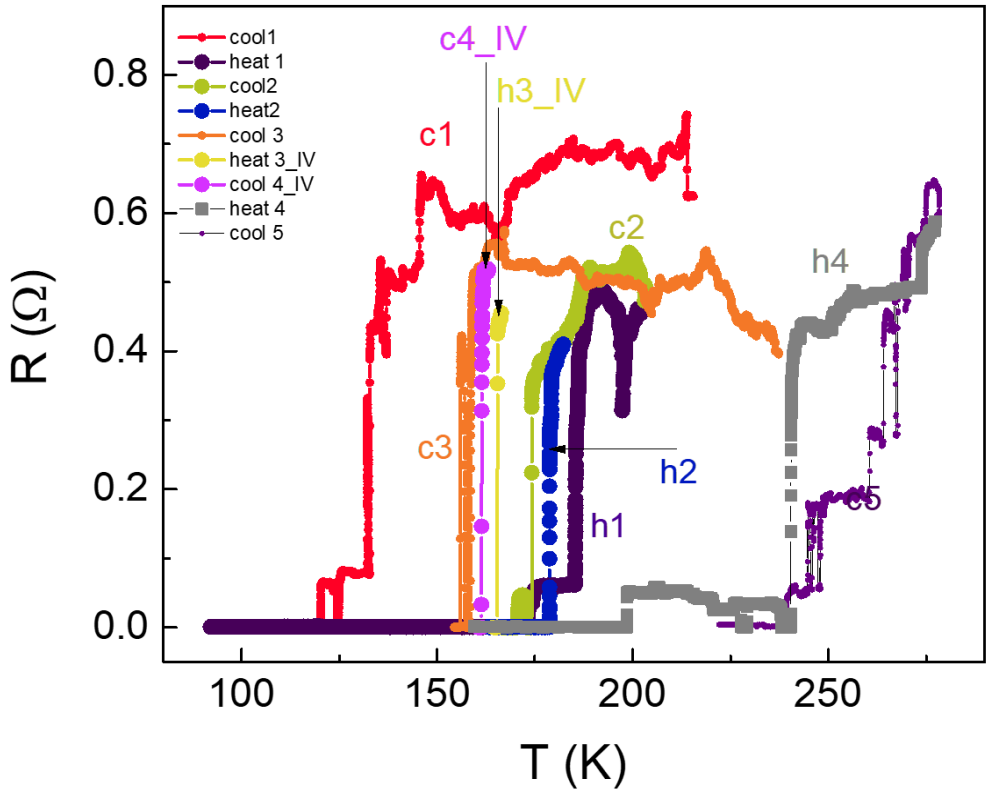

Fig. S1. Resistance vs temperature during all thermal cycles. Some of the cycles also show a re-entrant behaviour as seen previously. Here, "c(n)" denotes cooling during n[th] thermal cycle, while "h(n)" denotes heating.

**Section III: Resistance vs temperature in log-scale for first cooling**

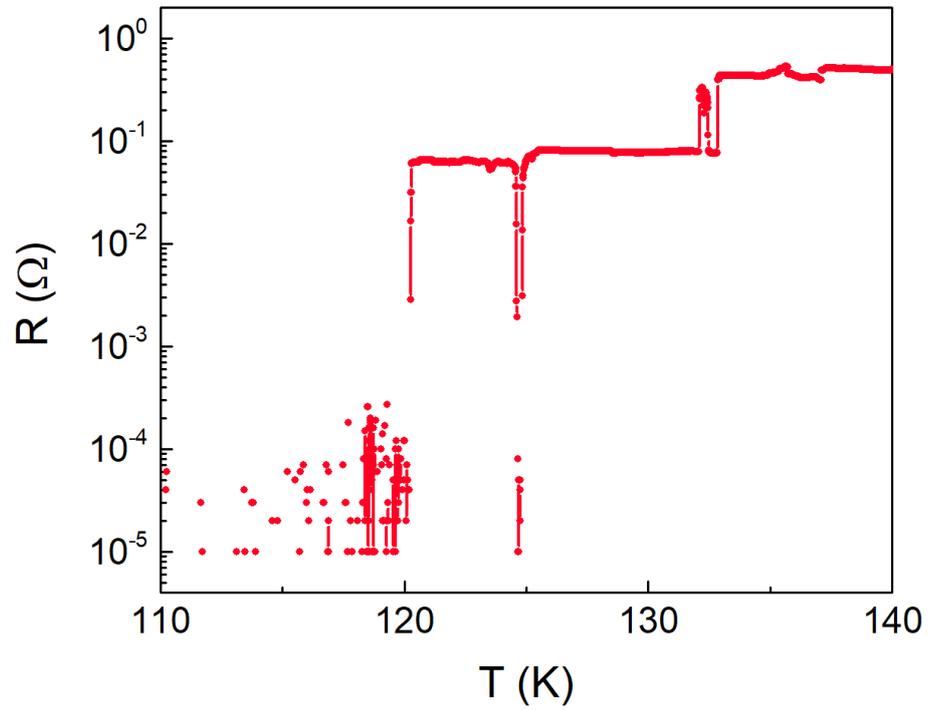

Fig. S2. Resistance vs temperature in log-scale for the first cooling cycle, showing that the resistance drops by almost five decades after the transition to ~10 µΩ.

**Section IV: Re-entrant transition in cooling cycle 3**

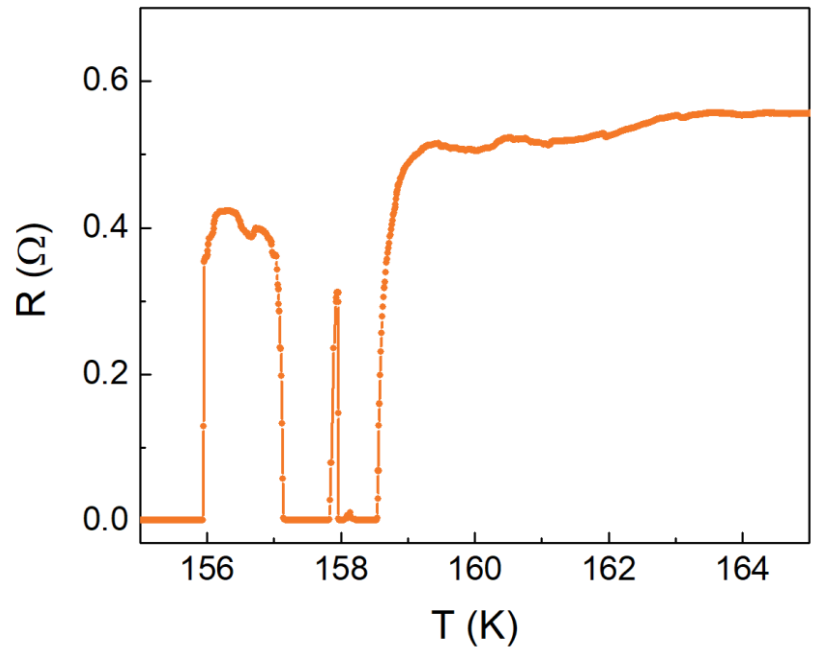

Fig. S3. Resistance vs temperature for third cooling cycle, displaying the re-entrant behaviour. This has also been previously observed in [1].

**Section V: Differential resistance $\frac{dV}{dI}$ at $T = 165.42$ K, $165.39$ K, and $165.36$ K**

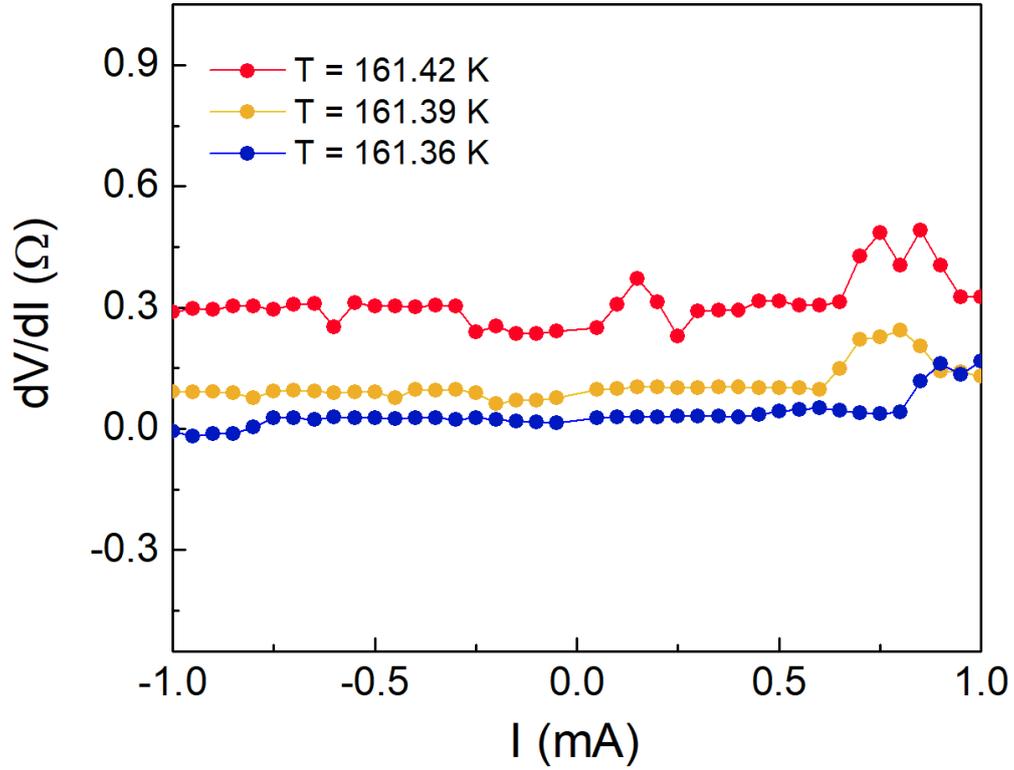

Fig. S4. The differential resistance $\frac{dV}{dI}$ from the measured $I-V$ characteristics at $T = 165.42$ K, $165.39$ K, and $165.36$ K during cooling cycle 4. The data does not exhibit any low-bias anomaly, and remains largely featureless.

**Section VI: Steps in $I-V$ characteristics at 161.39 K while cooling**

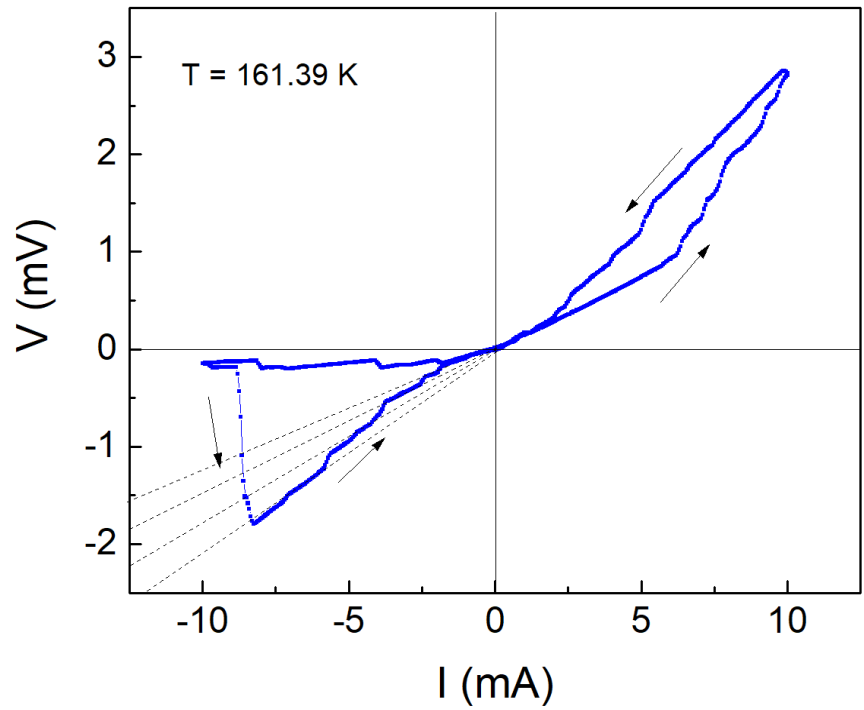

Fig. S5. Steps observed in the $I-V$ characteristics in both positive and negative $I$ sweep directions. While the positive sweep direction is presented in Fig. 3c of the main text, here we have indicated the steps obtained when the current was swept from -10 mA to 0 (dashed lines).

**Section VII: Steps in the $I-V$ characteristics while heating**

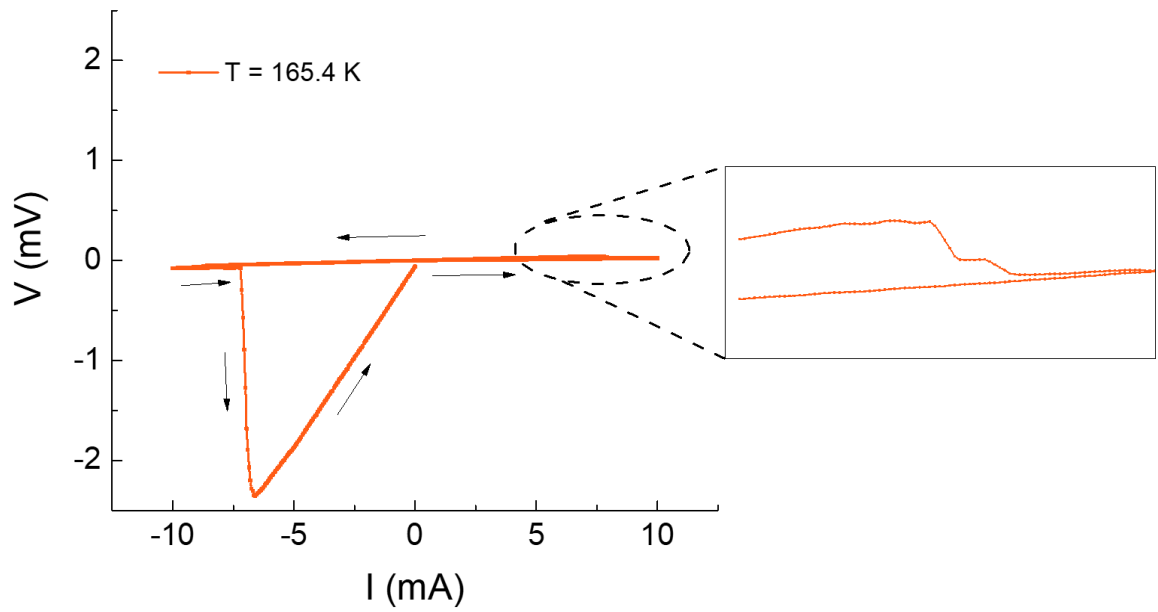

Fig. S6. Steps observed in the $I-V$ characteristics during the heating cycle at $T = 165.4$ K.

**Section VIII: $R - I$ characteristics while cooling at T = 161.39 K for both positive and negative currents**

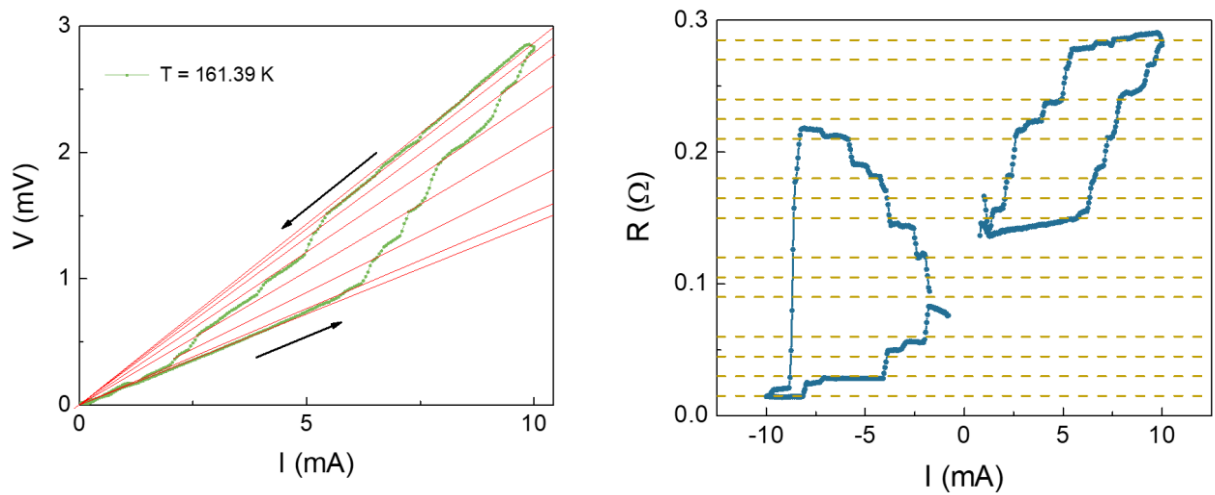

Fig. S7. (Left) Steps observed (for positive $I$) in the $I - V$ characteristics while cooling at $T = 161.39$ K. Note the absence of steps up to $I \sim 5$ mA, while the drive current was increased. Above this critical current the resistance increases rapidly in steps. (Right) $R$ (= $V/I$) obtained from the $I - V$ data for the entire cycle. The dashed lines are multiples of $R = 0.015$ Ω.

**Section IX: Hysteresis and stair-like pattern in the $I-V$ characteristics of device P20519FEE_20 (Ref [1])**

The hysteretic and step-like patterns in the $I-V$ characteristics of devices other than that presented in this paper. The data below was taken during the heating cycle with Device P20519FEE_20 whose detail can be found in Ref [1].

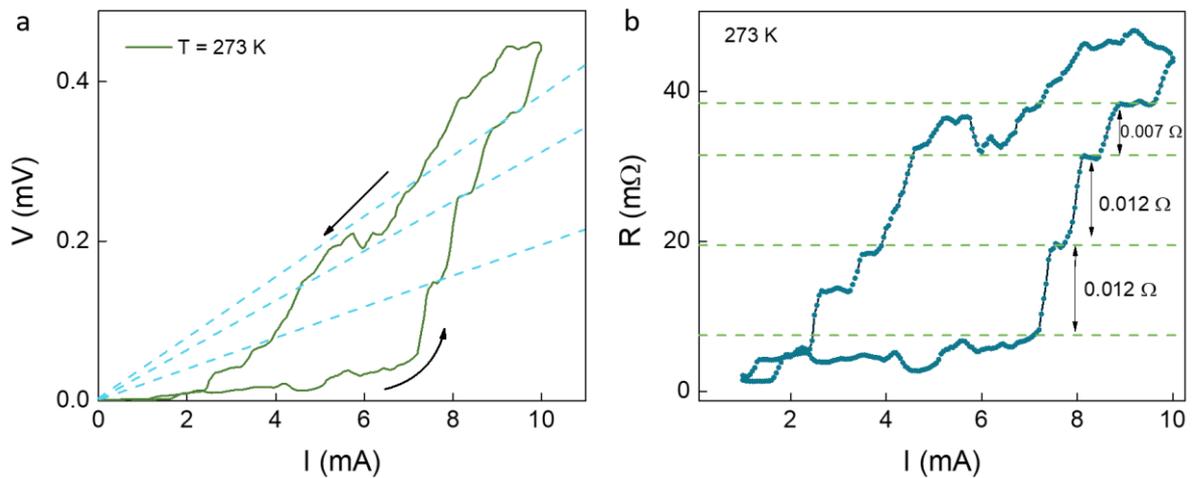

Fig. S8. (a) Steps observed in the $I-V$ characteristics in device P20519FEE_20 while the sample was heated through the resistive transition at $T = 273$ K [1]. The rapid increase in $V$ at $I = 7$ mA indicates the critical current behaviour. (b) $R \ (= V/I)$ obtained from the $I-V$ data, displaying step-like features. Note that the step-sizes in $R$ (~0.012 Ω and ~ 0.007 Ω) are in the same order as that observed in the current device.